Diffusive Charge Transport in Graphene on SiO$_2$

J.-H. Chen[1,2], C. Jang[2], M. Ishigami[2†], S. Xiao[2], E. D. Williams[1,2], and M. S. Fuhrer[1,2*]

[1]*Materials Research Science and Engineering Center, and* [2]*Center for Nanophysics and Advanced Materials, Department of Physics, University of Maryland, College Park, MD 20742 USA*

[†]*Present address: Department of Physics, University of Central Florida, 4000 Central Florida Boulevard, Orlando, Florida 32816-2385 USA*

*email: mfuhrer@umd.edu

We review our recent work on the physical mechanisms limiting the mobility of graphene on $SiO_2$. We have used intentional addition of charged scattering impurities and systematic variation of the dielectric environment to differentiate the effects of charged impurities and short-range scatterers. The results show that charged impurities indeed lead to a conductivity linear in density ($\sigma(n) \propto n$) in graphene, with a scattering magnitude that agrees quantitatively with theoretical estimates [1]; increased dielectric screening reduces scattering from charged impurities, but increases scattering from short-range scatterers [2]. We evaluate the effects of the corrugations (ripples) of graphene on $SiO_2$ on transport by measuring the height-height correlation function. The results show that the corrugations cannot mimic long-range (charged impurity) scattering effects, and have too small an amplitude-to-wavelength ratio to significantly affect the observed mobility via short-range scattering [3, 4]. Temperature-dependent measurements show that longitudinal acoustic phonons in graphene produce a resistivity linear in temperature and independent of carrier density [5]; at higher temperatures, polar optical phonons of the $SiO_2$ substrate give rise to an activated, carrier density-dependent resistivity [5].. Together the results paint a complete picture of charge carrier transport in graphene on $SiO_2$ in the diffusive regime.

# 1. Introduction

Charge carrier transport in graphene has been the focus of attention since the experimental realization of isolated graphene [6]. Much of the interest has arisen from the prospect of fabricating graphene into high speed electronic devices, which rely on the exceptional carrier mobility of the electronic material. However, to date, graphene devices fabricated on silicon dioxide substrate have shown field effect mobilities ranging from 0.1 to 2 m$^2$/Vs [7], much lower than the carrier mobility in its parent material (typically Kish graphite or highly ordered pyrolytic graphite,) which have mobilities close to 100 m$^2$/Vs at low temperature [8]. Understanding the scattering mechanisms that limit device performance is thus of vital importance. In this article, we review our recent work [1-3, 5] on the various possible scattering sources in graphene and their relative contributions to the conductivity σ:

$$\sigma^{-1} = \sigma_{ci}^{-1} + \sigma_{sr}^{-1} + \sigma_{mg}^{-1} + \sigma_{LA}^{-1} + \sigma_{PO}^{-1} + \sigma_{corr}^{-1} \qquad (1)$$

where the subscripts indicate the contributions due to charged impurities (ci), short-range scatterers (sr), midgap states (mg), longitudinal acoustic phonons (LA), polar optical phonons (PO) and surface corrugations (corr). The results reveal the path for improving the mobility of substrate-bound graphene.

A striking aspect of graphene charge transport, observed since the earliest studies, is the linear dependence of conductivity on charge carrier density $\sigma(n) \propto n$ over a wide range of carrier densities. Early theoretical work had predicted that in graphene the conductivity given by scattering by white-noise disorder $\sigma_{sr}$ [9], as well as for scattering by acoustic phonons $\sigma_{LA}$ [10] should be independent of carrier density.

It was soon pointed out that scattering by charged impurities [11-15] near the graphene sheet should produce a linear σ(n) of the form

$$\sigma_{ci}(n) = C_{ci} e \left| \frac{n}{n_{imp}} \right| \tag{2}$$

where $C_{ci}$ is a constant, $e$ the electronic charge and $n_{imp}$ the density of charged impurities. The linear $\sigma_{ci}(n)$ results from the $1/q$ dependence of the Coulomb potential on wavevector $q$; leading to a $1/k_F$ dependence of the scattering rate. A unique aspect of graphene, as opposed to other two-dimensional electron systems (2DES) is that the $1/k_F$ dependence is preserved even for the *screened* Coulomb potential in graphene [14], creating a clear dichotomy in graphene between long-range and short-range scattering potentials. Hwang, et al. [14] first calculated the screened Coulomb potential within the random phase approximation (RPA), and used the results to determine $C_{ci} \approx 5 \times 10^{15}$ V$^{-1}$s$^{-1}$. Novikov [16] noted that, beyond the Born approximation used in Ref. [14], an asymmetry in $C_{ci}$ for attractive vs. repulsive scattering (electron vs. hole carriers) is expected for Dirac fermions.

Other sources of scattering which could lead to a linear σ(n) have been proposed. Stauber et al. [17] proposed that mid-gap states associated with vacancies in graphene give rise to a conductivity of the form

$$\sigma_{mg}(n) = C_{midgap} e \left| \frac{n}{n_{imp}} \right| \left[ \ln\left(\sqrt{\pi n} R_0\right) \right]^2 \tag{3}$$

where $C_{midgap}$ is a constant and $R_0$ is the effective radius of the vacancy (on order the bond length in graphene). The logarithmic term leads to a slightly sub-linear dependence of conductivity on charge density.

Another proposal to explain the linear σ(n) has been the effect of geometric corrugation of graphene (i.e. "ripples"), present due to contact with a rough substrate [3] or as a result of thermally-activated out-of-plane motion of the graphene sheet [18, 19], or the presence of local modification of the bonding in graphene [20]. Katsnelson and Geim [19] have suggested that ripples in graphene produce a conductivity of the form

$$\sigma_{corr}(n) = C_{corr} e \left( \frac{n}{n_{imp}} \right)^{2H-1} \quad (4)$$

where $C_{corr}$ is a constant and the exponent $2H$ is given by the distance dependence of the height-height correlation function of a corrugated surface, e.g. $g(r) \propto r^{2H}$ at small $r$, where $g(r) = \left\langle (h(r_0 + r) - h(r_0))^2 \right\rangle$. In this scenario scattering by ripples could produce a linear σ(n) for $2H = 2$, a situation that would, in principle, occur for equilibrium fluctuations of a flexible membrane in a planar confining potential [21], or a constant σ(n) for $2H = 1$, typical of the much more common case of a non-equilibrium structure with short-range correlations [22].

Here we review our recent work on the nature of charge carrier scattering in graphene on $SiO_2$. By carefully changing the density of charged impurities on graphene, we prove that charged impurities indeed lead to the predicted linear $\sigma_{ci}(n) \propto n$ in graphene (Eq. (2)), with a scattering magnitude that agrees quantitatively with theoretical estimates [1]. By varying the dielectric environment of graphene, we vary the relative effects of two different limiters of the mobility, charged impurities ($\sigma_{ci}(n) \propto n$) and short-range scatterers ($\sigma_{sr}(n) \propto$ constant) [2]. In quantitative agreement with theoretical predictions, the scattering by charged impurities is *reduced* with increasing dielectric constant, due to the reduced charge carrier-impurity interaction; the scattering by short-range impurities is *increased*, due to reduced screening by the more weakly-interacting carriers.  The observation of an enhanced linear component of σ($n$) upon increase of dielectric constant rules out midgap states (Eq. (3)) as the source of the linear σ($n$) (see section 4 for more details). To evaluate the effect of corrugations, the topography of graphene on SiO$_2$ was measured, and the corrugations show a height-height correlation function which varies as $g(r) \propto r^{2H}$ with $2H = 1$; such correlations are expected to give rise to $\sigma_{corr}(n) \propto$ constant [3] according to Eq. (4).  Finally we demonstrate the effects of phonons on graphene on SiO$_2$ substrate.  The resistivity of graphene rises linearly with temperature at low temperature, with magnitude and lack of carrier density dependence in good agreement with theoretical predictions for acoustic phonon scattering [5].  At higher temperatures, an activated, carrier density-dependent resistivity arises due to scattering by polar optical phonons of the SiO$_2$ substrate [5], and becomes the dominant limiter of mobility above ~400K.  Together the results paint a complete picture of charge carrier transport in graphene on SiO$_2$ in the diffusive regime.

## 2. Experimental procedure

Graphene is obtained from Kish graphite by mechanical exfoliation [23] on 300nm SiO$_2$ over doped Si (back gate), with Au/Cr electrodes defined by electron-beam lithography.  Raman spectroscopy confirms that the samples are single layer graphene [24]. After fabrication, the devices are annealed in H$_2$/Ar at 300 °C for 1 hour to remove resist residues [1, 3], with additional bake-outs up to 490 K in ultra high vacuum (UHV) to remove residual absorbed gases before measurement.

To investigate the effects of charged impurities, dielectric screening and phonons, the devices are mounted on a liquid helium cooled cold finger in an UHV chamber with a heater so that the temperature of the device can be controlled from 16 K to 490 K. With the devices at low temperature, small quantities of potassium are deposited on graphene using a getter source. The potassium is subsequently removed by heating the devices to 490 K. Water vapor is introduced using a leak valve, and easily desorbs with annealing. Phonon effects are investigated by controlling the temperature of clean graphene devices and measuring the four-probe gate-dependent resistivity $\rho(V_g)$ of graphene *in situ*.

The graphene morphology was measured in a JEOL SPM system which incorporates scanning tunneling microscopy (STM), atomic force microscopy (AFM), and scanning electron microscopy (SEM) as well as charge transport measurement capability in a UHV environment. SEM and AFM are used to locate graphene devices on the insulating SiO$_2$ substrate, non-contact mode AFM is used to measure the corrugations of graphene as well as the surrounding SiO$_2$ substrate, and STM is used to measure the atomic structure of graphene.

## 3. Charged impurity scattering

To vary the density of charged impurities on graphene, a controlled potassium flux is deposited on the clean graphene device in sequential 2-second intervals at a sample temperature $T = 20$ K in UHV. Potassium donates one or part of its valence electron to graphene [25], and the resulting potassium ion acts as a charged impurity. The density of deposited K was varied from zero to ~$5 \times 10^{12}$ cm$^{-2}$, more than ten times the initial trapped charge density of SiO$_2$/graphene interface. This corresponds to potassium concentrations varying from zero to ~$1.6 \times 10^{-3}$ per C atom. The gate-voltage-dependent conductivity $\sigma(V_g)$ was measured *in situ* for the pristine device, and again after each doping interval. After several doping intervals, the device was annealed in UHV to 490 K to remove weakly adsorbed potassium [26], then cooled to 20 K and the doping experiment repeated; four such runs (Runs 1-4) were performed in total. As discussed below, some irreversible changes were observed during the first cycle of potassium deposition and removal, probably indicating irreversible reaction of potassium with defects or impurities in the graphene or on the underlying SiO$_2$ substrate. The subsequent deposition and removal cycles showed reversible behavior, indicating that the behavior observed in these runs is due to the variation in charged impurity concentration.

Fig. 1 shows the conductivity vs. gate voltage for the pristine [3] device and at three different doping concentrations at 20K in UHV for Run 3. For $V_g$ not too near $V_{g,min}$ and not too large (as sub-linearity might appear due to short-range scattering, see section 4 for details), the conductivity can be fit (see Fig. 1) by

$$\sigma(V_g) = \begin{cases} \mu_e c_g (V_g - V_{g,\min}) + \sigma_{res} & V_g > V_{g,\min} \\ -\mu_h c_g (V_g - V_{g,\min}) + \sigma_{res} & V_g < V_{g,\min} \end{cases} \quad (5)$$

where $\mu_e$ and $\mu_h$ are the electron and hole field-effect mobilities, and $c_g = 1.15 \times 10^{-4}$ F/m$^2$ is the gate capacitance per unit area for 300 nm thick SiO$_2$ substrate, and $\sigma_{res}$ is the residual conductivity which is determined by the fit. The minimum conductivity is discussed in more detail in Ref. [1]. The mobilities are reduced by an order of magnitude during each run, and recover upon annealing. The electron mobilities ranged from 0.081 to 1.32 m$^2$/Vs over the four runs, nearly covering the range of mobilities for graphene on SiO$_2$ substrate reported to date in the literature (~0.1 to 2 m$^2$/Vs) [7]. As K-dosing increases and mobility decreases, the linear behavior of $\sigma(V_g)$ (see Fig. 1) associated with charged impurity scattering dominates, as predicted theoretically [14]. For the clean graphene and at the lowest K-dosing level, sub-linear behavior is observed for large $|V_g - V_{g,\min}|$ as anticipated when short-range scattering is included; this sublinear behavior is examined further in section 4.

For uncorrelated scatterers, the mobility should depend inversely on the density of charged impurities, $1/\mu \propto n_{imp}$ (Eq. (2)). We assume $n_{imp}$ varies linearly with dosing time $t$ as potassium is added to the device. In Fig. 2 we plot $1/\mu_e$ and $1/\mu_h$ vs. $t$, which are linear, in agreement with $1/\mu \propto n_{imp}$, hence verifying that Eq. (2) describes charged impurity scattering in graphene. The inset to Fig. 2 shows that, although the $\mu_e$ and $\mu_h$ are not identical upon potassium dosing, their ratio is fairly constant at $\mu_e/\mu_h = 0.83 \pm 0.01$. Novikov [16] predicted $\mu_e/\mu_h = 0.37$ for an impurity charge $Z = 1$, however the asymmetry is expected to be reduced when screening by conduction electrons is included.

The concentration of potassium ions on the sample (of order $10^{-3}$ potassium ions/carbon atom) is difficult to determine directly. Therefore, to determine the magnitude of the charged impurity scattering (the constant $C_{ci}$ in Eq. (2)) we compare the shift in the minimum conductivity point, which probes charge donated by potassium to graphene, with the reduction in mobility. Fig. 3 shows $V_{g,min}$ as a function of $1/\mu_e$ which is proportional to impurity density. As discussed above, Run 1 differs from Runs 2-4 due to irreversible reaction of potassium on the first run. After Run 1, subsequent runs are very repeatable, other than an increasing rigid shift of the curves to more negative voltage. Adam, et al. [15] predicted that the minimum conductivity occurs at the added carrier density $\bar{n}$ at which the average impurity potential is zero, i.e. $\Delta V_{g,min} = -\bar{n}e/c_g$, where $\bar{n}$ is a function of $n_{imp}$, the impurity spacing $d$ from the graphene plane, and the dielectric constant of the $SiO_2$ substrate. The theoretical lines in Fig. 3 are given by the *exact result* of Adam et al.[15], and follow an approximate power-law behavior of $\Delta V_{g,min} \propto n_{imp}^b$ with $b = 1.2$~$1.3$, which agrees well with experiment. The only adjustable parameter is the impurity-graphene distance $d$; we show the results for $d = 0.3$ nm (a reasonable value for the distance of potassium on graphene [25, 27, 28]), and $d = 1.0$ nm (the value used by Adam, et al.). Since $\Delta V_{g,min}$ gives an independent estimate of $n_{imp}$, the quantitative agreement in Fig. 3 verifies that $C_{ci} \approx 5\times10^{15}$ $V^{-1}s^{-1}$ in Eq. (2), as expected theoretically.

## 4. Distinguishing short and long-range scattering by variable dielectric screening

The dielectric environment is expected to strongly affect the carriers confined in the one-atom-thick layer of carbon. The dielectric environment determines the effective fine structure constant in graphene, $\alpha = \dfrac{4\pi e^2}{(\kappa_1 + \kappa_2)hv_F}$, where $h$ is the Plank constant, $v_F$ the Fermi velocity and $\kappa_1$ and $\kappa_2$ the dielectric constant of the materials on the two sides of the graphene layer. Increasing ($\kappa_1+\kappa_2$) leads to a reduction in α, and reduces the Coulomb interaction of the carriers with charged impurities and the scattering from them. In contrast, the dielectric does not modify the atomic-scale potential of short-range scatterers, and there the leading effect is the reduction of screening by the charge carriers, which increases short range scattering resulting in lower high-density conductivity.

The opposing effects of increased dielectric screening on the strength of short-ranged and long-ranged disorders are demonstrated by measuring the changes of the four-probed conductivity σ($V_g$) of graphene upon deposition of water (ice) layers. A cleaned graphene device is cooled to 77K in UHV and deionized water was introduced through a leak valve attached to the chamber. The water gas pressure (determined by a residual gas analyzer) was $5 \pm 3 \times 10^{-8}$ torr. The amount of ice deposited was estimated by assuming a sticking coefficient of one and the ice layer density of $9.54 \times 10^{14}$ cm$^{-2}$ [29].

Fig. 4(a) shows σ($V_g$) for two different sample conditions, pristine graphene and graphene covered by 6 monolayers of ice. We assume that the conductivity is the sum of contributions from long-range and short-range scatterers, added in inverse according to Matthiessen's rule: $\sigma(n)^{-1} = \sigma_{ci}^{-1} + \sigma_{sr}^{-1}$. For $V_g$ not too near $V_{g,min}$, the conductivity can be fit (see Fig. 4(a)) by

$$\sigma(V_g)^{-1} = \begin{cases} \left(\mu_e c_g \left(V_g - V_{g,min}\right)\right)^{-1} + \sigma_{sr,e}^{-1} & V_g > V_{g,min} \\ -\left(\mu_h c_g \left(V_g - V_{g,min}\right)\right)^{-1} + \sigma_{sr,h}^{-1} & V_g < V_{g,min} \end{cases} \quad (6)$$

Here we discuss the symmetric part of the mobility, $\mu_{sym} = (\mu_e + \mu_h)/2$, and that of the conductivity from short-ranged scatterers, $\sigma_{sr,sym} = (\sigma_{sr,e} + \sigma_{sr,h})/2$. The antisymmetric contribution to the mobility results from corrections to scattering beyond RPA [16], and the antisymmetric contribution to the conductivity likely results from a contact effect [2, 30]. Figs. 4 (b) and 4(c) show $\mu_{sym}$ and $\sigma_{sr,sym}$ as a function of the number of ice layers. The mobility (Fig. 4(b)) of pristine graphene is 0.9 m$^2$/Vs, which is typical for clean graphene devices on SiO$_2$ substrates at low temperature. As the number of water layers increases, the mobility increases by over 30%, and saturates after about 3 layers of ice to about 1.2 m$^2$/Vs. In contrast, the conductivity due to short-range scatterers (Fig. 4(c)) decreases from 280 $e^2$/h to 170 $e^2$/h. The decrease in conductivity due to short-range scatterers shows a similar behavior as the mobility, saturating after several layers of ice have been added, suggesting they have the same origin. The absence of any sharp change in the conductivity or mobility at very low ice coverage rules out ice itself acting as a significant source of short- or long-range scattering. This is corroborated by the absence of a shift in the gate voltage of the minimum conductivity, consistent with physisorbed ice [29] not donating charge to graphene [1, 7, 15]. Furthermore, the fact that saturation of the dielectric behavior is observed after only a few layers indicates that: 1) the ice film is continuous, and has bulk dielectric properties after formation of a few monolayers; 2) the distance $d$ of charged impurity to graphene is small (less than 1 nm), so for an ice layer thicker than this distance, the bulk of the Coulomb impurity potential is affected by the dielectric and the screening effect saturates.

Quantitatively, by adding ice on top of the graphene, the effective fine structure constant, $\alpha$, is reduced from ~ 0.81 to ~ 0.56. We have calculated within RPA the change in the mobility and short-range conductivity upon changing $\alpha$ from 0.81 to 0.56; the results are shown as dashed lines in Fig. 4(b) and (c), which agree very well with the data.

The increase of slope of the linear component of $\sigma(n)$ of >30% upon addition of a dielectric layer is strong evidence that the linear $\sigma(n)$ in pristine samples arises from charged impurity scattering. Mid-gap states (Eq. (3)) in particular would show a decreased conductivity upon increasing dielectric constant, due to the atomic scale of the scattering potential. The nature of the scattering leading to $\sigma_{sr}$ is unclear. This scattering could results from weak atomic scale defects which are not in the unitary scattering limit considered for vacancies by Stauber et al.[17], or perhaps results from the corrugations of graphene on $SiO_2$ (discussed below in Section 5). The effect of varying dielectric constant on scattering from corrugations in graphene has not been studied thereotically.

## 5. Corrugation effects of graphene on $SiO_2$

In order to determine the role of corrugations in charge carrier scattering, the surface corrugation of graphene on $SiO_2$ was measured by non-contact mode AFM and STM in UHV. Fig. 5(a) shows an AFM image of graphene as well as the neighboring $SiO_2$ substrate. Fig. 5(b) shows the height-height correlation function [3] for the graphene and $SiO_2$ surface. Notably, graphene is smoother than the $SiO_2$ substrate, suggesting that the finite stiffness of graphene acts to smooth out corrugations. Both correlation functions rapidly increase as $g(r) \sim r^{2H}$ at short distances, with similar exponents $2H = 1.11 \pm 0.013$ for graphene and $2H = 1.17 \pm 0.014$ for $SiO_2$. A crossover at the correlation length and saturation at mean square roughness at large distances follow the short-distance behavior. As seen in the figure inset, interpolating the intersection of the power-law and saturated regimes yields values of the correlation length [31], which are $\xi = 32 \pm 1$ nm for graphene and $\xi = 23 \pm 0.6$ nm for $SiO_2$. The similar exponents and slightly larger correlation length of the graphene sheet is consistent with the graphene morphology being *determined* by the underlying $SiO_2$ substrate; the larger correlation length and smaller roughness of the graphene surface arise naturally due to the energy cost for out-of-plane deformation of graphene. The measured exponent of $2H \sim 1$ for graphene on $SiO_2$ indicates that corrugations of graphene on $SiO_2$ should result in a conductivity nearly independent of charge carrier density according to Eq. (4), therefore similar to short-ranged scattering [19], which may contribute to the carrier-density-independent term $\sigma_{sr}$ discussed in Section 4.

A quantitative evaluation of the impact of the ripples requires a realistic understanding of their structure and amplitude[4]. By dimensional analysis [32], the one-dimensional Fourier transform of the height-height correlation function, $G(r) = 2\sigma^2 (r/L_c)^{2H}$ will vary with wave-vector as $G(q) \sim q^{-(2H+1)}$, where σ is the rms roughness and $L_c$ is the correlation length. This yields a wave-vector dependence of the surface corrugations $A(q) = Aq^{-(2H+1)/2}$. The magnitude $A$ can be estimated by calculating its functional dependence on σ, $L_c$ and $2H$ [4].

For the observed short-range height correlations, e.g. $2H = 1.1$ of the graphene on $SiO_2$ [3], the amplitude of height corrugations varies linearly with wavelength. Given the measured correlation length and rms roughness of respectively $L_c = 32$ nm and $\sigma = 0.19$ nm, the ratio of amplitude to wavelength is $A(q)/\lambda \approx 0.007$. Resistivity due to scattering from such corrugations is expected to be proportional to $(A(q)/\lambda)^2$ [19, 33], and hence we expect that $\sigma_{corr}^{-1}$ is small in magnitude and unlikely to contribute significantly to the sum in Eq. (1).

## 6. Phonon scattering in graphene on $SiO_2$

We now turn to the temperature dependence of the resistivity ρ of graphene. Figs. 6(a) and 6(b) show $\rho(V_g, T)$ for two samples at seven different gate voltages plotted on a linear scale. The $\rho(V_g, T)$ curves are linear in temperature at low $T$ with a slope independent of carrier density (gate voltage), i.e. $\rho(V_g, T) = \rho_0(V_g) + \rho_A T$, with $\rho_A = (4.0 \pm 0.5) \times 10^{-6}$ $h/e^2 K$ as indicated by the short-dashed lines.

Acoustic phonon scattering is expected [10, 17, 34, 35] to give rise to a linear resistivity independent of carrier density, i.e.

$$\rho_{LA} = \left(\frac{h}{e^2}\right)\frac{\pi^2 D_A^2 k_B T}{2h^2 \rho_s v_s^2 v_F^2}, \tag{7}$$

where $k_B$ is the Boltzmann constant, $\rho_s = 7.6 \times 10^{-7}$ kg/m$^2$ is the 2D mass density of graphene, $v_F = 10^6$ m/s is the Fermi velocity, $v_s$ is the sound velocity, and $D_A$ the acoustic deformation potential. For LA phonons, $v_s = 2.1 \times 10^4$ m/s and our experimentally determined slope gives $D_A = 18 \pm 1$ eV, in good agreement with theoretical [34-38] and experimental [39, 40] expectations.

In contrast to the low-$T$ behavior, the resistivity at higher $T$ is highly non-linear in $T$, and becomes significantly dependent on $V_g$, increasing for decreasing $V_g$. An activated temperature dependence was predicted by Ref. [41] due to remote interfacial phonon (RIP) scattering[42] by polar optical phonons of the SiO$_2$ substrate. Following Ref. [41], we fit the data by adding an extra term $\rho_B(V_g,T)$ representing the RIP contribution to the resistivity:

$$\rho(V_g,T) = \rho_0(V_g) + \rho_{LA} + \rho_{PO}(V_g,T); \quad \rho_{PO}(V_g,T) = BV_g^{-\alpha}\left(\frac{1}{e^{(59 meV)/k_B T}-1} + \frac{6.5}{e^{(155 meV)/k_B T}-1}\right)$$

(8)

The form of the expression in parenthesis in $\rho_{PO}(V_g,T)$ is chosen to match surface polar optical phonons in SiO$_2$ with $\hbar\omega \approx 59$ meV and 155 meV, with a ratio of coupling to the electrons of 1:6.5 [41, 43]. Fig. 6(c) and 6(d) show a global fit to Eq. (8) (solid lines) to the data for two samples. In addition to the low-temperature resistivity $\rho_0$, and linear term $\rho_{LA}$ determined above, only two additional global parameters in Eq. (8) ($B = 0.607$ $(h/e^2)$V$^\alpha$ and $\alpha = 1.04$) are used to fit the seven curves each for two devices. The magnitude of the RIP scattering resistivity predicted by Fratini and Guinea [41] is on the order of a few $10^{-3}$ $h/e^2$ at 300 K, also in agreement with the observed magnitude. RIP results in a long-ranged potential, which gives rise to a density-dependent resistivity in graphene, similar to charged impurity scattering. Thus RIP naturally explains the magnitude, temperature dependence, and charge carrier density dependence of $\rho_{PO}(V_g,T)$, hence we consider RIP scattering to be the most likely origin of $\rho_{PO}(V_g,T)$.

## 7. Limits to mobility

Fig. 7 shows the temperature dependence of the mobility of Sample 1 and Sample 2 (same as in section 6) at $n = 10^{12}$ cm$^{-2}$ ($V_g = 14$ V), as well as the limits due to scattering by LA phonons, polar optical phonons of the SiO$_2$ substrate, and charged impurities. As shown in Fig. 7, even for the cleanest graphene devices fabricated to date, impurity scattering is the still the dominant factor limiting the mobility for $T <$ 400 K. For comparison, the temperature-dependent mobility in Kish graphite and pyrolytic graphite from ref. [8] are also shown; these are the two materials commonly used as sources for exfoliated graphene on SiO$_2$. The significantly higher mobility at low temperature in Kish and pyrolytic graphites compared to graphene is a strong indication that the impurity scattering in graphene on SiO$_2$ is not due to point defects present in the parent material, but rather is likely caused by charged impurities in the SiO$_2$ substrate [1, 15]. It is important to note that the closeness of the room-temperature mobility values for graphene and bulk graphite is a coincidence, and removing impurity scattering in graphene will greatly increase not only the low temperature mobility, but the room temperature mobility as well.

## 8. Conclusion

Our data give a complete picture of the current limitations and future promise of graphene as an electronic material. At the present state-of the art of materials preparation, the mobility of graphene on $SiO_2$ at low and room temperature is limited by charged impurity scattering, likely due to charged impurities in the $SiO_2$ substrate [1, 15], although the possibility that it may also be influenced by impurities deposited on graphene during the fabrication process cannot be ruled out. Above 200K, optical phonons from the $SiO_2$ substrate become an important limiting factor to the overall mobility. Corrugations in graphene on $SiO_2$ should produce a very small limiting resistivity that is independent of density[3, 19], and together with other short-ranged scatterers, have minor contribution to device resistivity. Increasing the low-temperature mobility of graphene can be accomplished by either (1) reducing the number of charged impurities, or (2) reducing their effect. The former has been demonstrated by removing the substrate altogether, followed by high current annealing of graphene to produce samples with mobility on order 200,000 $cm^2$/Vs at low temperature [44]. Reducing the effect of charged impurities can be accomplished by increasing the dielectric constant of the graphene environment; as shown above, a modest increase of the average dielectric constant from 2.45 to 3.55 resulted in an increase of mobility of over 30%[2]. However, increased dielectric constant will enhance scattering by short-ranged disorder; this results in a net decrease in conductivity for carrier densities greater than $3 \times 10^{12}$ cm$^{-2}$ in our experiments [2], and the crossover could occur at lower density for larger dielectric constant changes.

If charged impurity scattering can be reduced, the room-temperature mobility, limited by the extrinsic RIP scattering due to SiO$_2$ phonons could be improved to $4 \times 10^4$ cm$^2$/Vs, comparable to the best field-effect transistors [45]. With proper choice of substrate [46, 47], or by suspending graphene [44], the intrinsic limit of mobility of $2 \times 10^5$ cm$^2$/Vs at room temperature could be realized [5]. This would dramatically enhance the application of graphene field-effect devices to chemical sensing, high-speed analog electronics, and spintronics. In addition, ballistic transport over micron lengths would open the possibility of new electronic devices based on quantum transport operating at room temperature.

**Acknowledgements:** This work was supported by the U.S. ONR, an NRI-MRSEC supplemental grant, NSF-UMD-MRSEC grant DMR 05-20471 and NSF-DMR grant 08-04976. Useful discussions with S. Adam are gratefully acknowledged.

**Figures and captions**

Figure 1

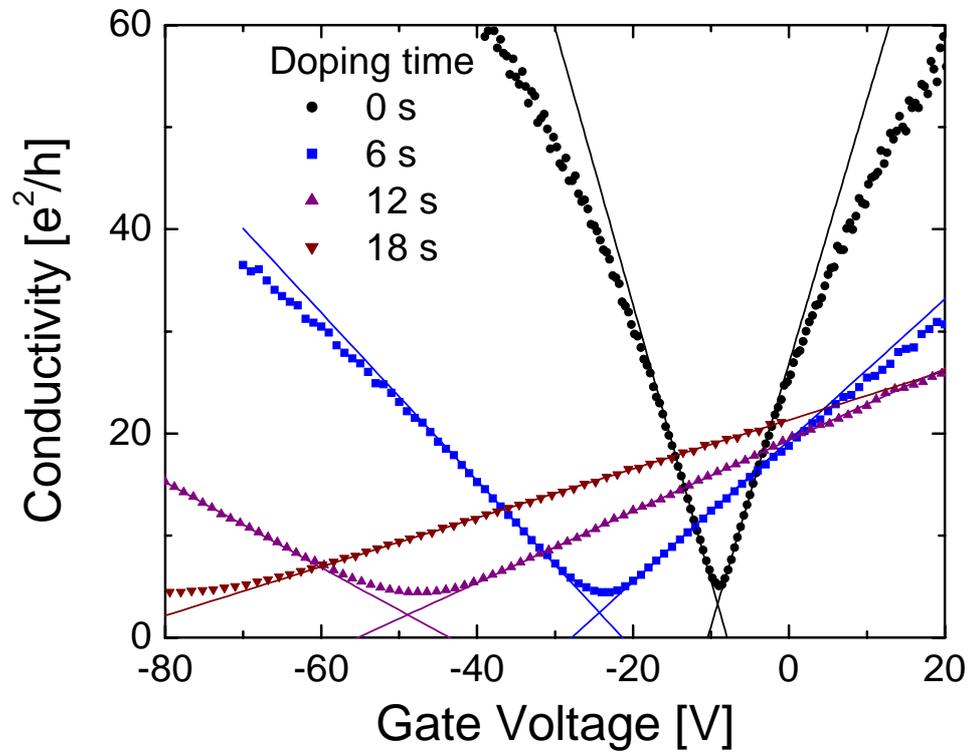

Fig. 1. The conductivity (σ) vs. gate voltage ($V_g$) curves for the pristine sample and three different doping concentrations taken at 20K in ultra high vacuum (UHV) are shown. The potassium dosing rate is $dn_{imp}/dt = (2.6$~$3.2)\times10^{15}$ m$^{-2}$s$^{-1}$. Data are from Run 3. Lines are fits to Eq. (5), and the crossing of the lines defines the points of the residual conductivity and the gate voltage at minimum conductivity ($\sigma_{res}$, $V_{g,min}$) for each data set.

Figure 2

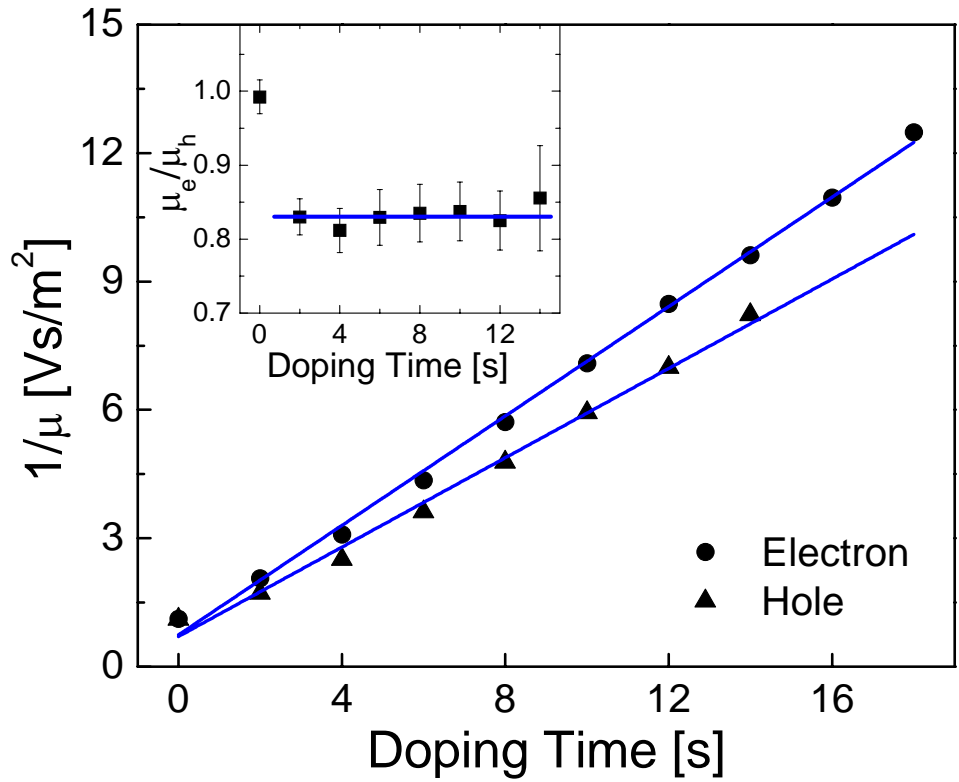

Fig. 2. Inverse of electron mobility $1/\mu_e$ and hole mobility $1/\mu_h$ vs. doping time. Lines are linear fits to all data points. Inset: The ratio of $\mu_e$ to $\mu_h$ vs. doping time. Error bars represent experimental error in determining the mobility ratio from the fitting procedure. Data are from run 3 (same as Fig. 1).

Figure 3

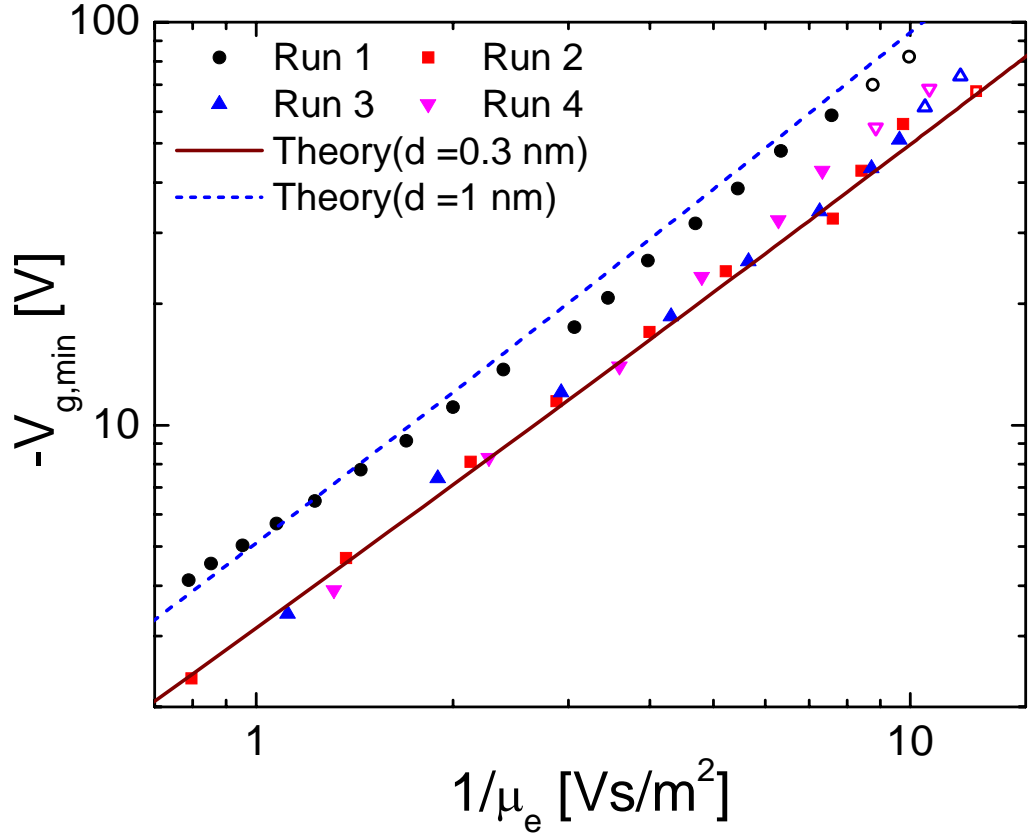

Fig. 3. The gate voltage of minimum conductivity $V_{g,min}$ is shown as a function of inverse mobility, which is proportional to the impurity concentration. All four experimental runs are shown. Each data set has been shifted by a constant offset in $V_{g,min}$ in order to make $V_{g,min}(1/\mu_e \rightarrow 0) = 0$, to account for any rigid threshold shift. The offset (in volts) is -10, 3.1, 5.6, and 8.2 for the four runs, respectively, with the variation likely to be due to accumulation of K in the SiO$_2$ on successive experiments. The open dots are $V_{g,min}$ obtained directly from the $\sigma(V_g)$ curves rather than fits to Eq. (5) because the linear regime of the hole side of these curves is not accessible due to heavy doping. The solid and short-dashed lines are from the theory of Adam et al.[15] for an impurity-graphene distance $d = 0.3$ nm (solid line) and $d = 1$ nm (short-dashed line), and approximately follow power laws with slopes 1.2 and 1.3, respectively.

Figure 4

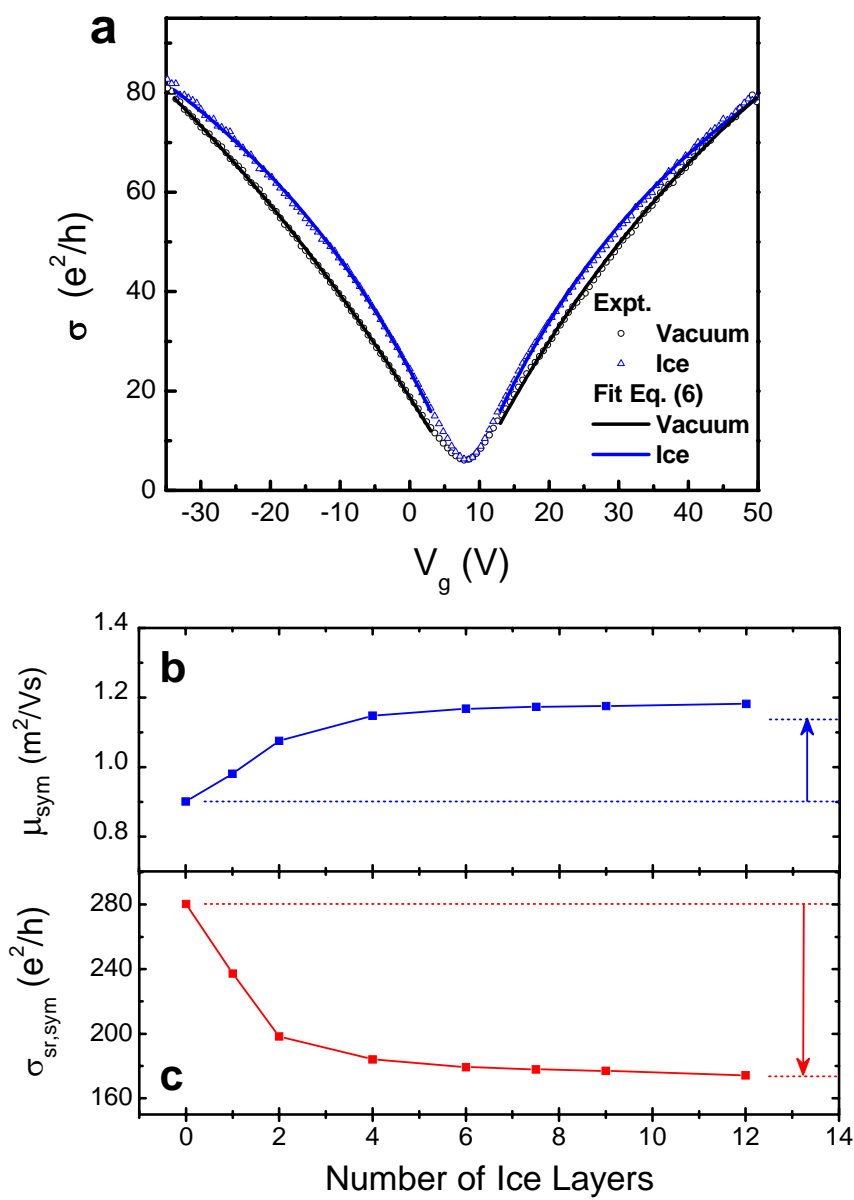

Fig. 4. (a) Conductivity of the graphene device as a function of back-gate voltage for pristine graphene (circles) and after deposition of 6 monolayers of ice (triangles). Lines are fits to Eq.. (6). (b) Symmetric part of the mobility $\mu_{sym}$ and (c) symmetric part of the conductivity from short-ranged scatterers $\sigma_{sr,sym}$ as a function of number of ice layers. Dashed lines show the values for pristine graphene and corresponding theoretical expectations for the ice-covered device.

Figure 5

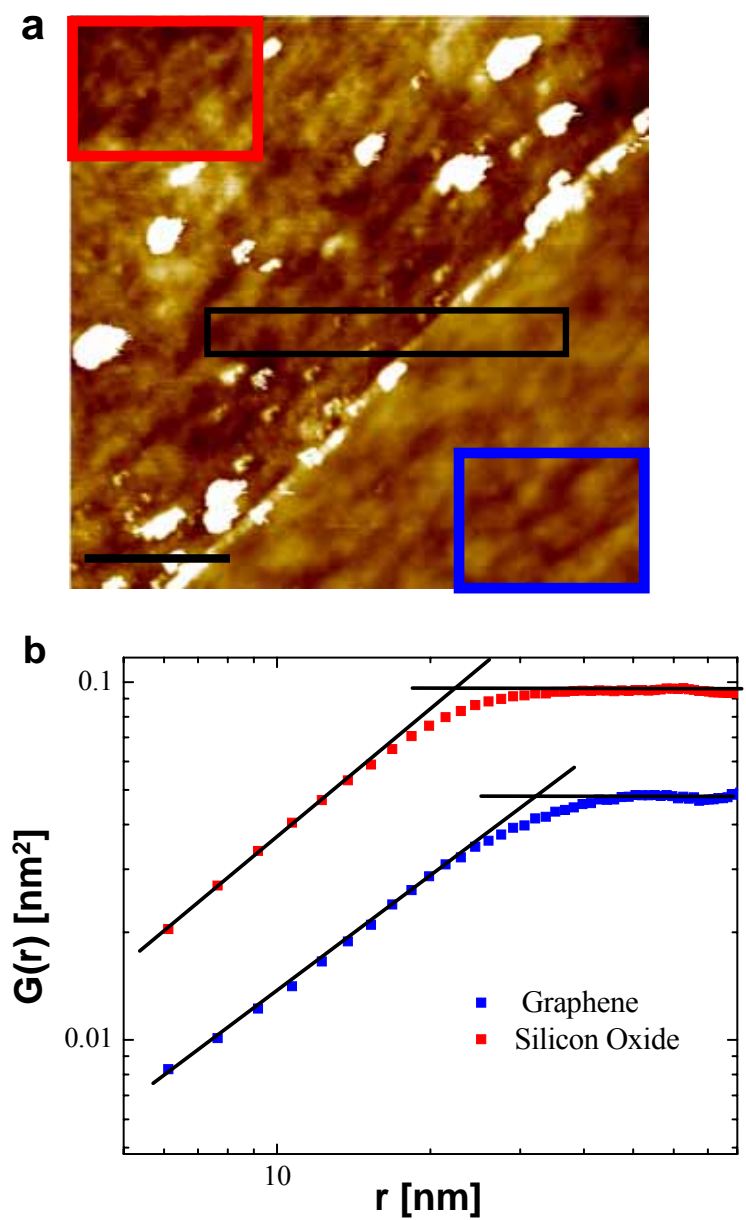

Fig. 5. (a) Non-contact mode AFM image, acquired in UHV, of a boundary between the graphene sheet and $SiO_2$ substrate. The graphene sheet occupies the lower right area of the image. The scale bar is 200 nm. (b) The height-height correlation function (G(r), see text) of the graphene sheet and $SiO_2$ surface. The lines are fits to the large and small length behaviors (power-law and constant, respectively), and the point of intersection indicates the correlation length. This analysis is performed by selecting data from Fig. 5(a), showing both graphene and $SiO_2$ surfaces in one scan, thus excluding the contribution of any tip-related artifact to the analysis.

Figure 6

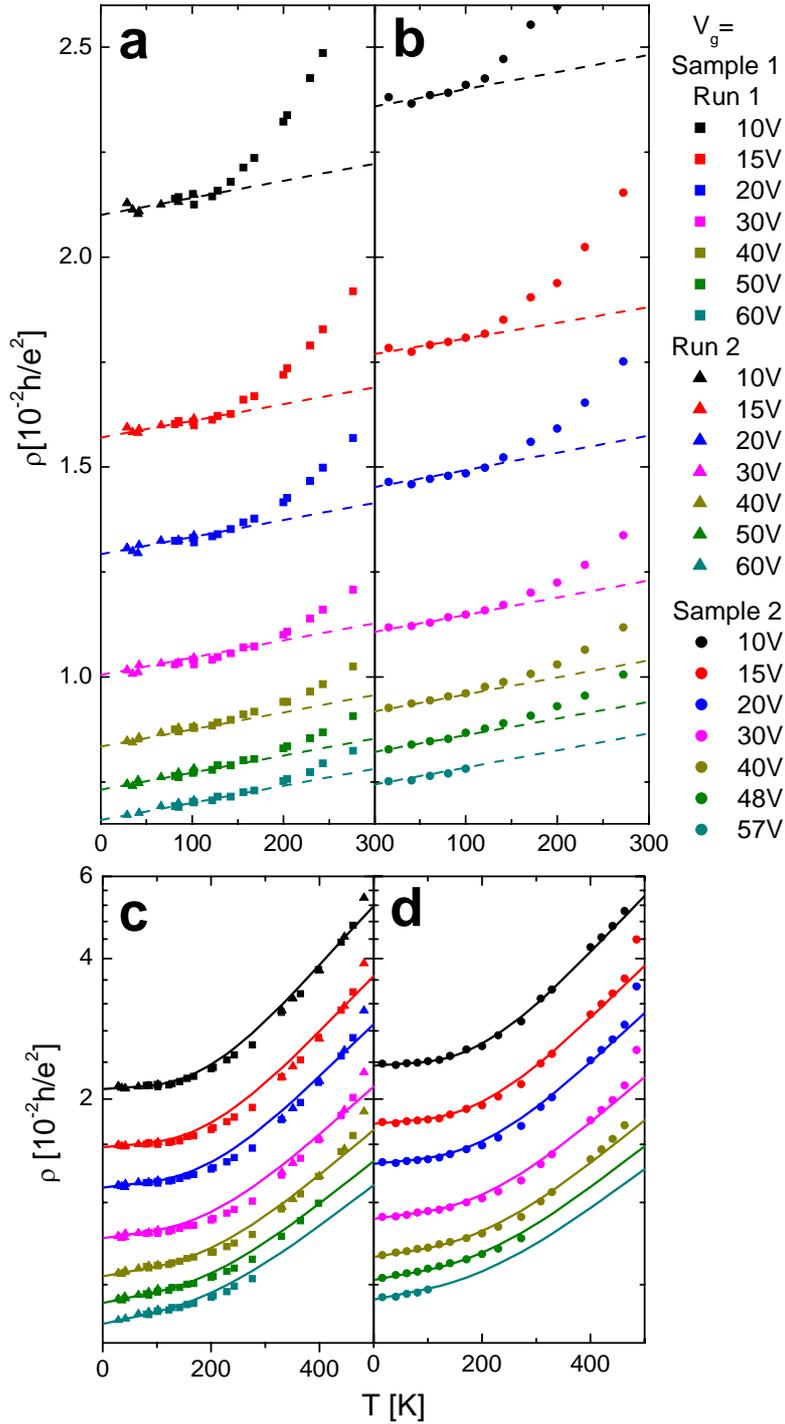

Fig. 6. (a) & (b), Resistivity of two graphene samples (Sample 1 (a) and Sample 2 (b)) as a function of temperature for gate voltages from 10 to 60 V. Short-dashed lines are fits to the linear T-dependence (Eq. (7)). (c) & (d), Same data as in Fig.6 (a) & (b) on a logarithmic scale. The solid lines are fits to Eq. (8) (acoustic phonon scattering in graphene plus polar optical phonon scattering due to the SiO2 substrate).

Figure 7

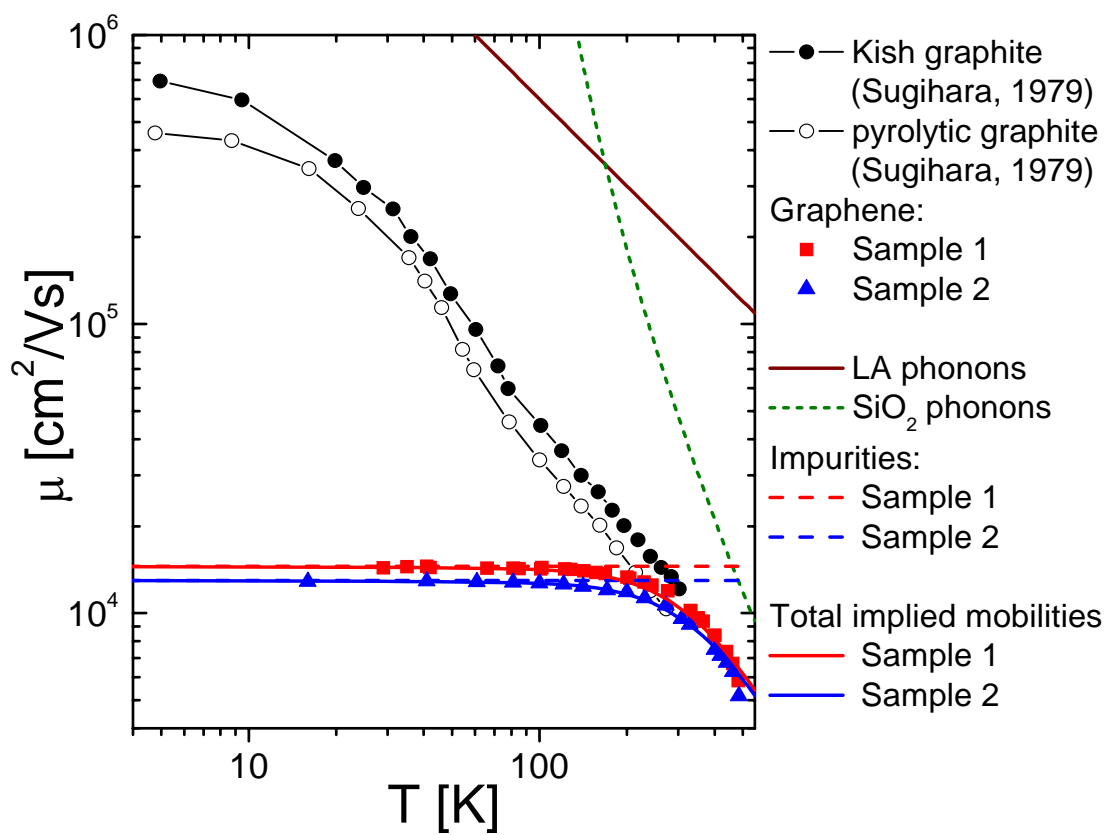

Fig. 7. The temperature-dependent mobilities of graphene Sample 1 (red squares) and Sample 2 (blue triangles) at $V_g$ = 14 V ($n$ = $10^{12}$cm$^{-2}$) are compared with Kish graphite (solid black circles) and pyrolytic graphite (open black circles) [8]. The mobility limits in graphene are shown for scattering by LA phonons (dark red solid line), remote interfacial phonon scattering (dark green short-dashed line), and impurity scattering (red and blue dashed lines). Red and blue solid lines show the expected net mobility for each sample, according to Matthiessen's rule.